\documentclass[twocolumn]{aastex63}

\usepackage{float}
\usepackage{amsmath}
\usepackage{afterpage}

\graphicspath{{./}{figures/}}

\newcommand{\HII}{\ion{H}{2}}
\newcommand{\NII}{[\ion{N}{2}]}

\newcommand{\SII}{[\ion{S}{2}]}

\newcommand{\OIII}{[\ion{O}{3}]}

\shorttitle{The PNLF with Superpositions}
\shortauthors{Chase et al.}

\begin{document}

\title{The Effect of Superpositions on the Planetary Nebula Luminosity Function}

\correspondingauthor{Robin Ciardullo}
\email{rbc@astro.psu.edu}

\author[0000-0002-0304-5701]{Owen Chase}
\affil{Department of Astronomy \& Astrophysics, The Pennsylvania
State University, University Park, PA 16802, USA}

\author[0000-0002-1328-0211]{Robin Ciardullo}
\affil{Department of Astronomy \& Astrophysics, The Pennsylvania
State University, University Park, PA 16802, USA}
\affil{Institute for Gravitation and the Cosmos, The Pennsylvania
State University, University Park, PA 16802, USA}

\author[0000-0003-2451-739X] {Martin M. Roth}
\affiliation{Leibniz Institute for Astrophysics Potsdam (AIP), An der Sternwarte 16, 14482 Potsdam, Germany}

\author[0000-0001-7970-0277]{George H. Jacoby}   
\affiliation{NSF’s NOIRLab, 950 N. Cherry Ave., Tucson, AZ 85719, USA}

\begin{abstract}
Planetary nebula (PN) surveys in systems beyond $\sim 10$~Mpc often find high-excitation, point-like sources with \OIII\ $\lambda 5007$ fluxes greater than the apparent bright-end cutoff of the planetary nebula luminosity function (PNLF\null).  Here we identify PN superpositions as one likely cause for the phenomenon and describe the proper procedures for deriving PNLF distances when object blends are a possibility.  We apply our technique to two objects:  a model Virgo-distance elliptical galaxy observed through a narrow-band interference filter, and the Fornax lenticular galaxy NGC\,1380 surveyed with the MUSE integral-field unit spectrograph.  Our analyses show that even when the most-likely distance to a galaxy is unaffected by the possible presence of PN superpositions, the resultant value will still be biased towards too small a distance due to the asymmetrical nature of the error bars.  We discuss the future of the PNLF in an era where current ground-based instrumentation can push the technique to distances beyond
$\sim 35$~Mpc.
\end{abstract}


\accepted{to ApJ}
  
\section{Introduction}
\label{sec:intro}
One of the most disconcerting aspects of modern cosmology is the current ``tension'' between the Hubble constant inferred from a $\Lambda$CDM analysis of the microwave background and that derived from local measurements of the extragalactic distance scale.  If one applies the simple physics of sound-wave propagation to the early universe, then the power spectrum of the microwave background, combined with the assumption of a Cosmological constant, yields a value for the modern day Hubble constant of $67.37 \pm 0.54$~km\,s$^{-1}$\,Mpc$^{-1}$ \citep{Planck+20}.  Yet measurements of $H_0$ derived from the Cepheid calibration of SN~Ia are significantly larger than this, $H_0 = 73.2 \pm 1.3$ and $73.04 \pm 0.04$~km\,s$^{-1}$\,Mpc$^{-1}$ \citep{Riess+21, Riess+22}, as are estimates based on megamasers \citep{Pesce+20} and gravitationally lensed quasars \citep{Wong+20}. Meanwhile, a SN~Ia calibration based on the tip of the red giant branch (TRGB) lies in-between the two regimes \citep[$H_0 = 69.6 \pm 0.8$ (statistical) $\pm 1.9$ (systematic) km\,s$^{-1}$\,Mpc$^{-1}$;][]{Freedman+20}, while an analysis of Year 1 data from the Dark Energy Survey favors the lower Planck value \citep{Park+20}.  Given this situation and the known susceptibility of Hubble Constant measurements to various types of biases and systematic error \citep{Shah+21, Lopez_22},  additional methods for determining precise distances throughout the local universe are needed.

One possible technique, which was the subject of intense investigation a quarter century ago, is the \OIII\ $\lambda 5007$ planetary nebula luminosity function (PNLF\null).  Although the bright-end cutoff of the PNLF lacks a robust theoretical explanation \citep[see the discussion in][]{Ciardullo_12}, empirical evidence from galaxies within $\sim 10$~Mpc indicates that it is an exceptionally reliable standard candle.  PN studies in galaxies as diverse as the large Sb spiral M31 \citep{Merrett+06, Bhattacharya+19}, the bulgeless small-spiral M33 \citep{Ciardullo+04}, the giant interacting lenticular NGC\,5128 \citep{Hui+93}, and the normal elliptical NGC\,4494 \citep{Jacoby+96}, have been unable to detect any change in the absolute luminosity of the PNLF cutoff with galactocentric radius \citep[but see][]{Sambhus+06}. Similarly, observations of galaxies in groups and clusters, such as Triangulum \citep{Ciardullo+91}, the M81 Group \citep{Jacoby+89, Ciardullo+02a}, Leo~I \citep{Ciardullo+89b, Feldmeier+97, Ciardullo+02a}, Virgo \citep{Jacoby+90}, and Fornax \citep{Spriggs+21} almost always place all the galaxies comfortably within a typical group diameter of $\sim 1$~Mpc.  (The lone exception is in Virgo, where the PNLF clearly resolves the M84/M86 system which has long been known to be infalling into Virgo from behind \citep{Binggeli+93, Bohringer+94}.)  Finally, external tests against distances from Cepheids and the TRGB for galaxies within $\sim 10$~Mpc show excellent agreement, as does a comparison to the megamaser distance of NGC\,4258 \citep{Ciardullo+02a, Ciardullo_12}.  All this evidence suggests that careful PNLF measurements should be capable of producing extragalactic distances to a precision comparable to that of Cepheid variables and the TRGB\null.  The technique would then serve as an important cross-check for both techniques and unite the Pop I and Pop II distance ladders.

Despite these positive results, PNLF observations declined over the past two decades.  There were two principal reasons for this.  The first was due to technology: with the telescopes and instrumentation of the era, precise PNLF distances beyond $\sim 15$\,Mpc were difficult to obtain.  This, along with  our poor understanding of the physics of the PNLF cutoff and the difficulty associated with measuring absolute PN luminosities in the Milky Way \citep[e.g.,][]{Chornay+20, Chornay+21}, meant that if one wanted to use the PNLF to measure the Hubble constant, it would have to be a three-step process, involving a Cepheid (or TRGB) calibration of the PNLF in nearby galaxies and a PNLF calibration of Type~Ia supernovae.  Since the range of the PNLF technique was not that much greater than that for Cepheid/TRGB observations, the additional step associated with a PNLF calibration added an unnecessary uncertainty into the distance ladder.

The more serious issue affecting the PNLF was the discovery of an unknown systematic error associated with the method.  As mentioned above, for galaxies closer than $\sim 10$~Mpc, a direct comparison of PNLF distances with the distances obtained from the Cepheid and TRGB methods yields excellent results:  the scatter between the techniques is consistent with the internal errors of the methods, and there is no evidence of any systematic trend \citep{Ciardullo_12}.  However, beyond this threshold, a problem occurs.  As first noted by \citet{Ferrarese+00}, the PNLF distances to the elliptical galaxies of Virgo and Fornax are $\sim 0.2$~mag shorter than the Cepheid distances to the clusters' spirals, and this same offset is reflected in a comparison between the results of the PNLF and the Surface Brightness Fluctuation method  \citep[SBF;][]{Tonry+01}. This offset prompted \citet{Freedman+01} to exclude the PNLF from their final \textit{HST} Key Project Distance Ladder.

The sudden onset of the PNLF's systematic error at $D > 10$~Mpc limits the possible causes of the phenomenon.  \citet{Ciardullo+02a} argued that the discrepancy between the PNLF and SBF distances could be due to a small difference between the internal extinction in the nearby calibration galaxies (which are spirals) and the more distant targets of the PNLF and SBF methods (elliptical and lenticular systems).  But this conjecture did not explain the offset from the Virgo and Fornax Cepheid observations.  Alternatively, the problem may simply be due to the inclusion of interlopers in the PN samples.  At fainter magnitudes, it can be difficult to distinguish PNe from other emission-line objects without expensive spectroscopy, and, at magnitudes fainter than $m_{5007} \sim 25.5$, $z \sim 3$ Ly$\alpha$ emitting galaxies can be a serious source of contamination \citep[e.g.,][]{Ciardullo+02b, Gronwall+07}.  Yet another possibility, at least for galaxies in clusters, is contamination by foreground intergalactic planetary nebulae \citep[e.g.,][]{Feldmeier+04, Doherty+09, Longobardi+13}, and it is also possible that the shape of the PNLF in the brightest, most metal-rich cluster elliptical galaxies is different from that seen locally. Finally, since the spatial resolution of PN surveys is finite, observations in the most distant galaxies may suffer from object blending. If this were to happen, the PNLF could appear distorted, and a fit to the function may lead to an incorrect distance \citep{Jacoby+90}.

Until recently, a systematic investigation into all the possible causes of the PNLF's failure beyond $\sim 10$~Mpc was prohibitively expensive. But data from the VLT's Multi Unit Spectroscopic Explorer \citep[MUSE;][]{Bacon+10} are changing this.  Because MUSE is an integral-field unit (IFU) spectrograph, its spectra can be used to immediately remove interloping emission-line objects from a PN sample, thus eliminating one important source of the PNLF's systematic error.  Moreover, a wide range of galaxies have already been observed with MUSE, allowing the analysis of both spiral and elliptical galaxies in and out of galaxy clusters. 

MUSE observations have a number of other advantages for PNLF studies, including excellent image quality \citep[e.g.,][]{Fusco+20}, an effective bandpass that is $\sim 10$ times narrower than that delivered by most interference filters, and a spectral resolution high enough to resolve the velocity distribution of PNe within a galaxy.  These properties, when coupled with the VLT's large aperture, enable the PNLF technique to be applied to galaxies beyond $\sim 35$\,Mpc \citep{Roth+21}.  This is significantly further than the current range of the Cepheid and TRGB techniques, and effectively doubles the number of SN~Ia host galaxies within reach.  Moreover, at $\sim 35$~Mpc, the peculiar velocities of galaxies are small enough so that perturbations on the Hubble flow are less than 10\%.  If the PNLF's systematic error can be identified, then just a dozen measurements in a set of carefully selected galaxies could yield a $\sim 4\%$ estimate of the Hubble constant that is independent of the calibration of Type~Ia supernovae. 

A number of PNLF studies have been performed with MUSE \citep[e.g.,][]{Spriggs+21, Scheuermann+22}, including the very careful, pathfinding study of \citet{Roth+21}.  These analyses simply fit the observed distribution of PN apparent \OIII\ $\lambda 5007$ magnitudes to a truncated power law function \citep[e.g.,]{Ciardullo+89a, Mendez+01, Longobardi+13}.  The difference between the apparent magnitude of the cutoff, $m^*$, and the cutoff's assumed absolute magnitude, $M^*$, is the galaxy's apparent distance modulus.  However, this procedure is only valid if the flux of each source comes from a single PN\null.  One result of the \citet{Roth+21} study is the realization that the occurrence of superposed PNe, i.e., two planetary nebulae lying within a single element of spatial resolution, is not as uncommon as one might expect.  For example, in their analysis of the PNe within three MUSE fields in the Fornax lenticular galaxy NGC\,1380, \citet{Roth+21} identified 15 objects composed of two separate overlapping images with slightly different line-of-sight velocities. These data were taken in $0\farcs 75$ seeing and had a velocity resolution of $\sim 150$~km~s$^{-1}$.  Since the interference-filter surveys of past decades typically had no better than $1\arcsec$ seeing and no velocity resolution, it is reasonable to assume that at least some of the bright PNe found in these studies were actually superpositions of two separate objects.  If so, then the combined \OIII\ $\lambda 5007$ flux from the two PNe can create a source that is brighter than the nominal PNLF cutoff and/or distort the shape of the galaxy's intrinsic PN luminosity function.  The result would be a PNLF fit that underestimates the true distance to the galaxy.  Since the likelihood of this happening is proportional to the square of the distance, the effect may be at least partially responsible for the PNLF's systematic error in clusters such as Virgo. 

The possible effect of PN superpositions on the PNLF method was first examined by \citet{Jacoby+90} using a simple Monte Carlo simulation for a model elliptical galaxy.  These authors concluded that random PN superpositions could not explain all the ``overluminous'' PNe found in their survey of the Virgo cluster, but their analysis ended there; no quantitative analysis was performed on the effect of undetected blends on a PNLF distance determination.  Here we investigate the effect in detail and discuss how to interpret an observed PNLF which may include superposed objects.  In \S\ref{sec:theory}, we derive the expected form of the PNLF in the presence of blends and show how one should derive a PNLF distance when the existence of superpositions is a possibility.  In \S\ref{sec:narrow-band_results}, we examine the effect that PN superpositions have on narrow-band surveys of a typical Virgo-distance elliptical galaxy, and show that analyses that do not model the phenomenon will underestimate the true distance to the galaxy by $\gtrsim 0.1$~mag between 5 and 10\% of the time.  In \S\ref{sec:muse_parameters}, we turn our attention to PNLF surveys with MUSE and examine how the hyperparameters of seeing and spectral resolution translate into our ability to identify marginally-resolved pairs of PNe\null.  We then re-analyze the PNLF of the central field of NGC\,1380 using a Bayesian analysis which includes he possibility of blends.  We conclude by discussing PN superpositions in the context of other effects which could compromised a PNLF distance, and point out the next step needed to use the PNLF for Hubble constant studies.

\section{Fitting the PNLF with Blends}
\label{sec:theory}

The \OIII\ $\lambda 5007$ emission-line fluxes of planetary nebulae are typically expressed in magnitudes, with 
\begin{equation}
    m_{5007} = -2.5 \log F_{5007} - 13.74
\label{eq:magnitude}
\end{equation}
where the line flux is given in units of ergs~cm$^{-2}$\,s$^{-1}$.  For a PNLF consisting of only single (non-blended) objects, the brightest PNe in a large galaxy typically have absolute magnitudes near $M_{5007} \sim -4.5$ \citep{Ciardullo_12}, which is equivalent to $\sim 620 L_{\odot}$ of monochromatic light.  A PNLF containing superposed objects can theoretically produce sources that are up to 
0.75~mag brighter than this value, thereby skewing any measurement of the PNLF cutoff.

To model the PNLF in the case where PN superpositions may exist, we can use the fact that the spatial distribution of PNe within a galaxy is usually not that different from that of the system's light \citep[e.g.,][]{Jacoby+90, Hui+93, Merrett+06}. This allows us to write an expression for the likelihood of a PN blend as a function of the $V$-band surface brightness ($\mu_V$) at any position within a galaxy. This equation involves four parameters:

$\bullet$ The expected number of PNe brighter than some absolute \OIII\ magnitude, $M_x$, normalized to the bolometric luminosity of the underlying galaxy population.  (In other words, the number of PNe with $M_{5007} < x$ per solar luminosity of galaxy light.) This parameter, which is commonly called $\alpha_x$,  has been measured in more than two dozen spiral and elliptical galaxies, and for values of $M_x < -2$, generally does not vary by more than a factor of $\sim 2$.  Very roughly, all systems produce one~PN brighter than $M_{5007} = -2$ for every $\sim 5 \times 10^7 L_{\odot}$ ($M_{\rm bol} \sim -14.5$) of bolometric light \citep{Ciardullo+05, Ciardullo_10}.  Values of $\alpha_x$ for absolute magnitudes fainter than $M_{5007} = -2$ must be extrapolated using some assumed shape of the luminosity function.

$\bullet$ The target population's distance, $d$.  Obviously, the greater the distance to a galaxy, the more stars are present in a single spatial resolution element, and the greater the possibility that two PNe will be co-located on that element.

$\bullet$ The angular resolution of the data, $\theta$.  The effect of $\theta$ is similar to that of distance:  the poorer the image quality, the more luminosity is contained within a single resolution element, and the greater the chance that two distinct PNe will be present on the same element.  In practice, $\theta$ is not as simple as a single constant, as one's ability to resolve two, closely-separated emission-line sources is a complex function of several variables, including the relative brightness of the objects being observed and the signal-to-noise of the detection.  However, at its core, $\theta$ is proportional to seeing, and all other variables enter as second-order effects.  (This is especially true for a PNLF analysis since the most important superpositions are those where the two objects are bright and have comparable luminosities.)  Thus, for a first-order solution, a single value for $\theta$ is sufficient.

$\bullet$ The spectral resolution of the data expressed in terms of $f_v$, the fraction of superposed PNe that cannot be resolved via an analysis of the observed line profile. For surveys with interference filters, where the spectral resolution of the data is much lower than that needed to measure a system's line-of-sight velocity dispersion, this parameter is unimportant, as there is no way to disentangle the emission from two PNe falling onto the same spatial element.  In this limiting case, $f_v = 1$. For data taken with integral field unit spectrographs, a blended \OIII\ source can be decomposed into separate objects \textit{if} the radial velocities of the PNe are different enough to allow their individual emission lines to be resolved. If we approximate the line-of-sight velocity distribution of the stars in a galaxy as a Gaussian, then the calculation of $f_v$ is straightforward.  If $\Delta v$ is the minimum velocity separation that can be resolved by an IFU spectrograph, then the fraction of spatial superpositions that will remain spectrally unresolved is
\begin{equation}
f_v = {\rm erf} \left( \frac{\Delta v}{2\sigma_{\rm gal}} \right) 
\label{eq:fblend}
\end{equation}
where $\sigma_{\rm gal}$ is the galaxy's line-of-sight velocity dispersion at the location of the PN.

As is the case with $\theta$, the use of a single value for $\Delta v$ is an oversimplification, as one's ability to recognize an \OIII\ source as a blend of two objects depends on the signal-to-noise of the detection and the contrast in flux between the sources. Moreover \citet{Roth+21} showed that in some cases, PN superpositions can be identified by slight changes in the line profile as a function of spatial position in the data cube, i.e., $f_v$ can depend on $\theta$.  Nevertheless, for a first order analysis of the PNLF, a single mean value of $f_v$ can be used.

If we put this together with the definition of bolometric magnitude, then the expectation value for the number of unresolved PNe brighter than absolute magnitude $M_x$ co-located within a single resolution element of surface brightness $\mu_V$ is
\begin{equation}
\begin{split}
\lambda \simeq 0.01715  & \left( \frac{\alpha_x} {20 \times 10^{-9} \textrm{ PNe}\ L_{\odot}^{-1}} \right) 
\left( \frac{\theta}{1\arcsec} \right)^2 f_v
\\ &\hskip25pt \left( \frac{d}{10~{\rm Mpc}} \right)^2  10^{(22 - \mu_V) / 2.5}
\end{split}
\label{eq:lambda}
\end{equation}
where we have assumed a bolometric correction for the stellar population of B.C. $= -0.85$ \citep{Buzzoni+06}. Note that  $\theta$ and $f_v$ should be known from the properties of the instrument and the observing conditions, and $\mu_V$ can either be measured directly from the data or obtained from independent surface photometry.   Thus, $\lambda$ is a function of only two variables: $d$ and $\alpha_x$. 

Once we have the expectation value for the number of PNe within a resolution element, we can compute the likelihood that $n$ objects appear in that element.  Under Poisson statistics, this is 
\begin{equation}
p(n,\lambda) = \frac{\lambda^n e^{-\lambda}}{n!}
\label{eq:poisson}
\end{equation}
Thus, for any assumed galaxy distance and $\alpha_x$, the probability that an \OIII\ source is composed of $n$ separate objects is 
\begin{equation}
    w_n = {p(n,\lambda) \Bigg/ \displaystyle\sum_{i=1}^\infty p(i,\lambda)}
    \label{eq:weight}
\end{equation}
where the denominator does not automatically sum to one, since the $n=0$ term is omitted from the expression.   Thus, the values of $w_n$ can be thought of as a series of weights representing the relative likelihood of a given object being a PN superposition.  In the halo of a galaxy where the surface density of PNe is expected to be extremely low, $w_1 \simeq 1$ and all other weights will effectively be zero.  But if a galaxy $\sim 20$~Mpc away were observed with a $1\arcsec$ resolution element, then for $\alpha = 20 \times 10^{-9}$~PNe\,$L_{\odot}^{-1}$ and $f_v = 1$, an \OIII\ $\lambda 5007$ source projected at a location with $\mu_V = 20.8$~mag~arcsec$^{-2}$, would have a $\sim 10\%$ chance of being composed of multiple superposed PNe.

Next, we consider the luminosity function of the blended sources.  We start by letting $\phi_1(M) dM$ be the probability distribution function (PDF) of observing a single PN with absolute \OIII\ $\lambda 5007$ magnitude between $M$ and $M + dM$ over the full range of magnitudes, i.e., from $[-\infty, +\infty]$.  If we convert this function to linear units, $\phi_1(L) dL$, we can write down the probability of observing a single PN with luminosity between $L_1$ and $L_1 + dL$. If a second PN is superposed upon this first object and if the two fluxes can be treated as independent variables, then the probability of the combined source having luminosity $L_T = L_1 + L_2$ is
\begin{equation}
    \phi_2(L_T) = \phi_1(L) \ast \phi_1(L) = 
    \int_{-\infty}^\infty \phi_1(x) \phi_1(L_T - x) dx 
    \label{eq:convolution}
\end{equation}
In other words, the probability distribution function for the sum of two PN luminosities is simply the convolution of the normalized PNLF with itself.  Once this new function is calculated, it can then be translated back into magnitude space to give $\phi_2(M) dM$, the probability distribution versus magnitude for two superposed sources.

The above equation can be generalized to handle multiple superpositions, as the PDF for the combined flux of $n$ superposed objects is simply the convolution of that for $n-1$ objects with the single object distribution, $\phi_1(L) dL$. In theory, one can then compute a series of probability distributions, $\phi_i(M) dM$, each representing the magnitude distribution for a different number of superpositions.  In practice, however, these higher-order terms are not necessary: since PNe are relatively rare, superpositions of three or more objects can be ignored in all but the highest surface brightness regions of a galaxy, where faint-object photometry is likely to be unreliable.  Thus, the summation of equation~(\ref{eq:weight}) usually has just two terms:  one for single objects and one for two-object blends.

We note that in performing the above convolution, it is not necessary to extend the lower limit of the integral to zero.  For a superposition to increase the flux of a PN by more than 1\%, the fainter component must be within 5~magnitudes of the primary.  In practice, this means that one never needs to extend an integration more than $\sim 5$~mag fainter than the least luminous object in the sample.  More importantly, a more precise calculation of equation~(\ref{eq:convolution}) does not necessarily equate to a more accurate PNLF analysis, as the shape of the PN luminosity function at faint magnitudes is not universal.  For example, the PNLFs of star-forming systems have been observed to have multiple inflection points at fainter magnitudes, and the locations of these dips vary from galaxy to galaxy \citep[e.g.,][]{Jacoby+02, Reid+10, Ciardullo_10, Gonzalez-Rodriguez+15}.  Moreover, while the PN luminosity functions of older populations are generally monotonic, they do not always have the same faint-end slope, nor are they necessarily featureless \citep{Longobardi+13, Hartke+20, Bhattacharya+19, Bhattacharya+21}.   Fortunately, since the most important blends for PNLF distance measurements are those formed from the superposition of PNe in the brightest $\sim 1$~mag of the luminosity function, the errors introduced by not knowing the relative numbers of faint sources are minor at best.

\begin{figure}
\centering
\includegraphics[width=0.473\textwidth]{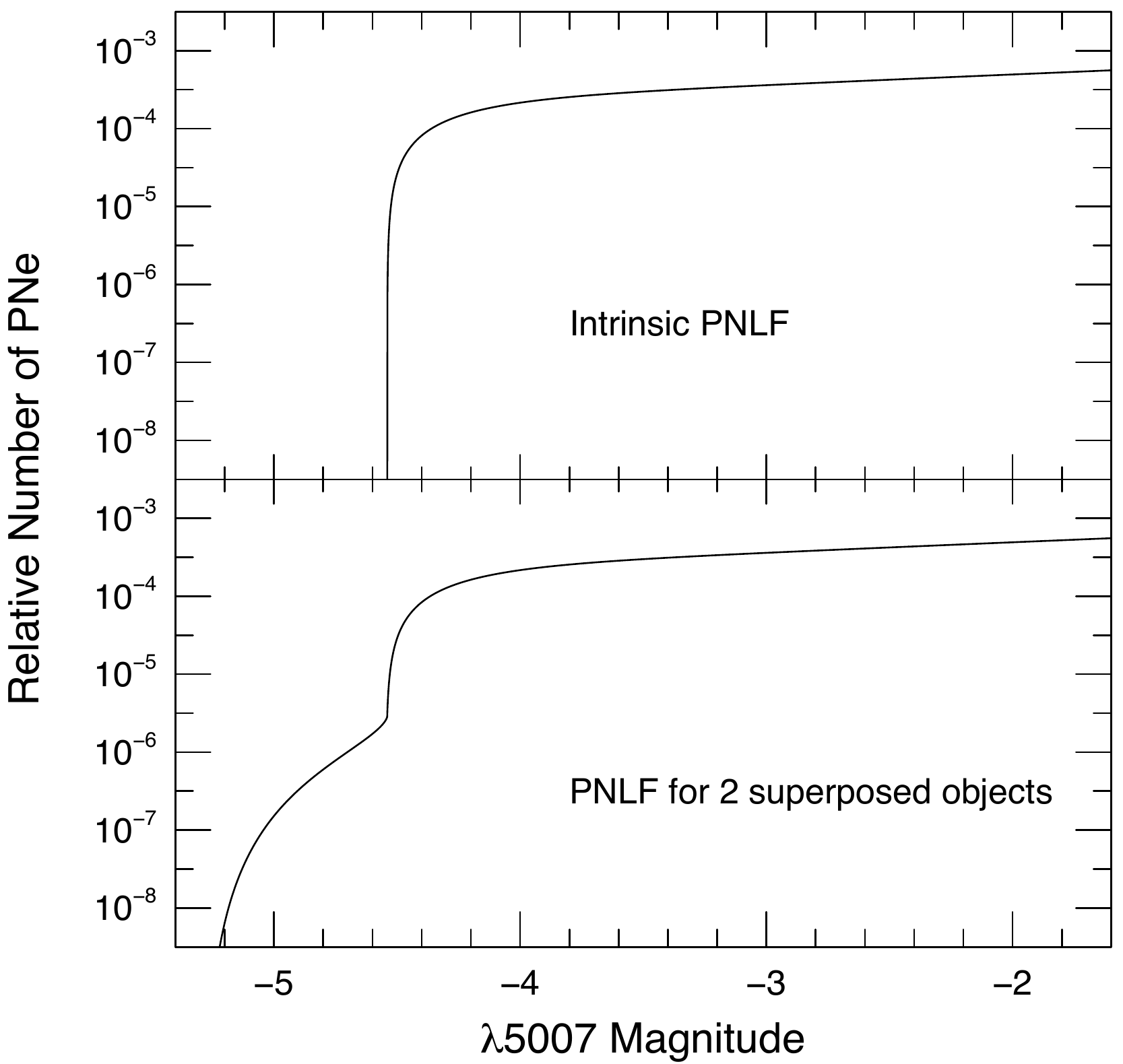}
\caption{\textit{Top Panel:}  The assumed shape of the single-object planetary nebula luminosity function, $\phi_1(M)$ from equation~(\ref{eq:pnlf}).
\textit{ Bottom Panel:} The shape of PDF  produced by two superposed planetaries, $\phi_2(M)$.  At any position in a galaxy, the expected PNLF will be a linear combination of these two functions.}    
\label{fig:pnlf_blend}
\end{figure}

Figure~\ref{fig:pnlf_blend} compares the PDF produced by two superposed sources to that formed by single objects.  The top panel of the plot is a single-object function, $\phi_1(M)$, which we have modeled via
\begin{equation}
\phi_1(M) \propto e^{0.307 M} \{ 1 - e^{3 (M^* - M)} \}
\label{eq:pnlf}
\end{equation}
with $M^* = -4.53$ \citep{Ciardullo+89a, Ciardullo_22}, though other forms of this PNLF are possible \citep[e.g.,][]{Longobardi+13, Gonzalez-Rodriguez+15, Bhattacharya+19, Valenzuela+19}.  The lower panel of Figure~\ref{fig:pnlf_blend} displays $\phi_2(M)$, the distribution function expected for two blended objects, as computed from equation~(\ref{eq:convolution}). Note that $\phi_2(M)$ does not go to zero at $M^*$, but instead assigns a finite likelihood to objects as luminous as $M = M^* - 0.75$.  Interestingly, the shape of this extension is not too dissimilar to that of the extreme bright-end of $\phi_1(M)$.  The effect of this coincidence will be discussed below.

We now have all the pieces required to measure the relative likelihood of an observed PNLF as a function of distance and $\alpha_x$.  First, at any position in a galaxy, we can use equation~(\ref{eq:weight}) and the galaxy's known surface brightness to estimate the relative probability that an \OIII\ source is composed of $n$ objects, rather than a single PN\null.  The expected probability distribution function for a PN at that position is then
\begin{equation}
    \phi^T(M) = \sum_{i=1}^\infty w_i \, \phi_i(M) 
    \label{eq:superposition}
\end{equation}
where for almost all applications, the summation includes only the first two terms of the series.  

Next we consider the effect of photometric errors on this expression.  If $G(m)$ represents the uncertainty of a PN measurement as a function of apparent magnitude $m$, then the observed distribution of PN magnitudes will be the convolution of the true distribution with the error function, i.e.,
\begin{equation}
    \phi^T_{\rm obs}(M) = \phi^T(M) \ast G(M + \mu) 
    \label{eq:errors}
\end{equation}
where $\mu$ is the galaxy's apparent distance modulus for the redshifted wavelength of \OIII\ $\lambda 5007$.  Typically, $G(m)$ is modeled as a Gaussian with a dispersion $\sigma(m)$ and a normalization of 1; in that case, the convolution captures only the effect of photometric errors.  However, it is possible to use an asymmetric kernel with a normalization less than one to simultaneously mimic the effects of both photometric errors and incompleteness.  For PNLF measurements, this is generally not necessary, since one usually does not need to press down to the faintest magnitudes where incompleteness is important.

We now let $m_{\rm lim}$ be the magnitude below which a PNe could not be detected (or, at least, where the detections become too incomplete for analysis).  If we normalize $\phi^T_{\rm obs}(M)$ to one between in the interval $[-\infty, m_{\rm lim} - \mu]$, then the likelihood of observing an \OIII\ source with apparent magnitude $m$ is simply $\phi^T_{\rm obs}(m-\mu)$ and the log likelihood of observing a set of $N$ PNe in a galaxy is
\begin{equation}
    \ln P = \sum_{i=1}^N \ln \phi^T_{{\rm obs},i} 
    \label{eq:probability}
\end{equation}

Finally, we can put a prior on $\alpha_x$ using the total amount of bolometric galaxy luminosity encompassed by the PN survey and the number of \OIII\ sources detected.  The latter is related to the total number of PNe by
\begin{equation}
    N_{\rm PN} = \sum_{i=1}^N w_{1,i} + 2 w_{2,i} + 3 w_{3,i} + \dots
\label{eq:npn}
\end{equation}
while the former is simply
\begin{equation}
    L_{\rm bol} = 10^{(M_{\rm bol}({\odot}) - V_{\rm tot} - {\rm B.C.} + \mu)/2.5} 
\label{eq:lbol}
\end{equation}
where for most stellar populations, the bolometric correction is B.C. $= -0.85$ \citep{Buzzoni+06}.  Through these equations, a given $\alpha_x$ yields a prediction for the total PN population, which, in turn, translates into a prior on $\alpha_x$ through the Poissonian distribution of equation~(\ref{eq:poisson}).  The likelihoods produced by this prior multiplied by those of  equation~(\ref{eq:probability}), yield a relative probability for any combination of distance and $\alpha_x$.  The most likely distance can then be found by marginalizing over $\alpha_x$.

As equation~(\ref{eq:npn}) shows, the prior on $\alpha_x$ depends on the value of $\lambda$ for each \OIII\ source, and those $\lambda$ values themselves depend on $\alpha_x$.  Thus, the computation of a PNLF likelihood is formally an iterative procedure, where one first estimates $\alpha_x$ using the number of \OIII\ sources detected in the survey and then computes a set of modified $\lambda$ values using equation~(\ref{eq:npn}).  In practice, however, such an approach is not necessary, as the vast majority of \OIII\ sources found in an extragalactic PN survey are expected to be single objects.  Consequently, to within a couple of percent, $N_{PN} = N_{\rm [O~III]}$ and the effect of equation~(\ref{eq:npn}) on the array of $\lambda$ values is minimal.  

It should be emphasized that the Bayesian scheme described above is necessitated by the fact that precise PNLF distances require at least $\sim 30$ to 50 objects in the top $\sim 1$~mag of the luminosity function \citep{Jacoby_97, Roth+21}.  To collect such a sample, one often has to survey the inner regions of a galaxy, and in those high surface-brightness areas, PN superpositions can and do occur.  Since the likelihood of a superposition ($\lambda$) depends on the surface density of stars, the expected luminosity function, $\phi^T(M)$, the photometric uncertainty, $G(m)$, and the limiting magnitude, $m_{\rm lim}$, will be different for every object.  Thus, the concept of fitting a single PNLF to all the \OIII\ sources in a galaxy is incorrect; instead the expected luminosity function should be tailored to each individual object. While the effect is small in nearby systems, its importance grows as the square of the distance, so that by $\sim 20$~Mpc, its effect cannot be ignored.  Thus, to obtain a high precision distance, histograms and/or cumulative distributions of PN magnitudes should only be employed for visualization purposes, as their use for quantitative analyses will likely lead to an erroneous result or an underestimated uncertainty.

\section{PNLF Measurements with Narrow-Band Imaging}
\label{sec:narrow-band_results}
To quantify the effect that the above analysis has on PNLF distances, we begin by considering the traditional method of conducting extragalactic PN surveys through narrow-band imaging.  Throughout the 1990s and early 2000's, most PNLF distances were obtained through the use of an interference filter, whose central wavelength was tuned to \OIII\ $\lambda 5007$ at the redshift of the target galaxy (see the summaries in \citet{Ciardullo+02a} and \citet{Ciardullo_06}).  Even today, narrow-band imaging has its uses:  while IFU spectrographs such as MUSE and VIRUS \citep{Hill+21} have their advantages (see \S\ref{sec:intro}), their fields-of-view are measured in arcminutes, and multiple pointings are generally needed to observe one galaxy.  In contrast, \OIII\ filters on wide-field imagers such as HyperSuprime Cam \citep{Miyazaki+18} and DECam \citep{Flaugher+15} can observe $\sim 2$ to 3~deg$^2$ at once, allowing PNe surveys to be conducted over entire galaxy clusters.  Though imaging surveys cannot approach the depths attainable by IFU spectroscopy, their greater entendue enables an entirely different class of science.

The full-width-half-maximum (FWHM) of most available narrow-band filters is typically between 30 and 75~\AA; these values are wide enough to capture the full velocity dispersion of a galaxy's stars and also avoid a degradation of the filter's performance due to the fast optics of a telescope. With filters of this width, there is no possibility of spectrally resolving two PNe co-located on the same spatial resolution element, meaning $f_v = 1$.  Thus, under poor-seeing conditions, we might expect the results for distant galaxies to be compromised by the effects of image blending.

\begin{figure*}
\includegraphics[width=18cm]{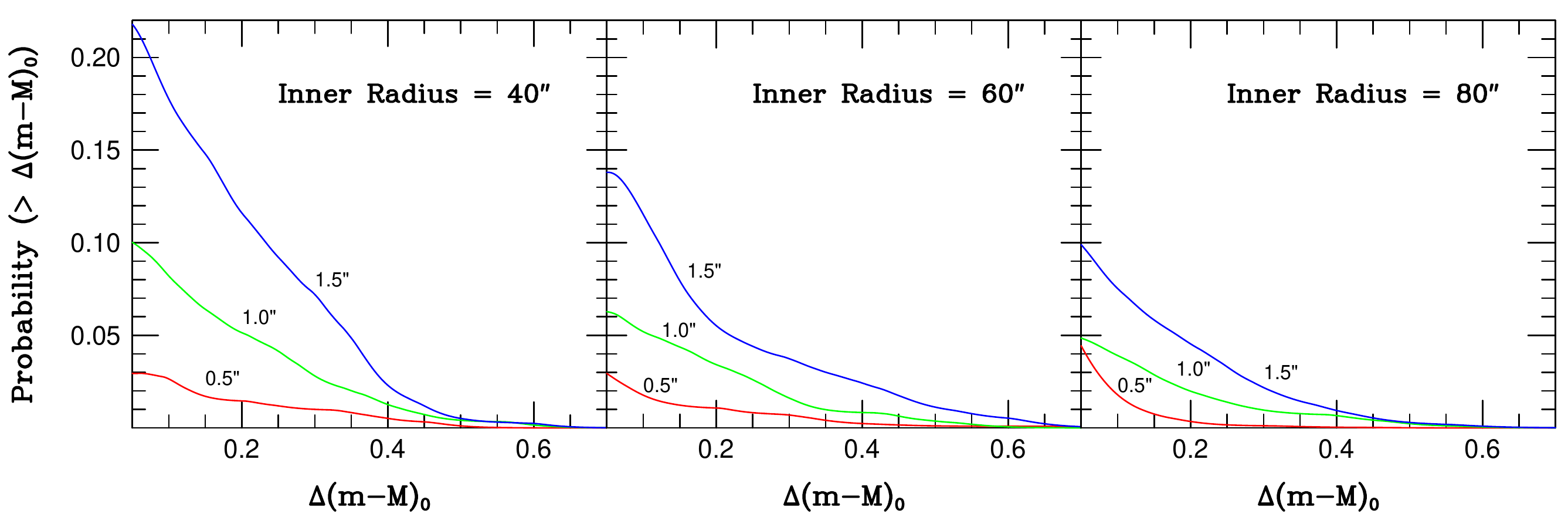}
\caption{The probability of deriving a distance modulus at least $\Delta (m-M)_0$ smaller than the true modulus by not taking PN blends into account, for a model elliptical galaxy with $(m-M)_0 = 31.25$.  The red, green, and blue curves represent simulations for $0\farcs 5$, $1\farcs 0$, and $1\farcs 5$ seeing, and the bright inner regions of the galaxy have been excluded as labeled in the panels.  Unless the image quality is excellent, there is a $\sim 5$ to 10\% likelihood that a PNLF survey with interference filters will underestimate the distance to a Virgo or Fornax elliptical galaxy by $\sim 0.1$~mag due to image blends. 
}
\label{fig:blend_simulation}
\end{figure*}

To investigate this possibility, we created an idealized giant elliptical galaxy using the $g$-band surface photometry of NGC\,4406 as a guide \citep{Cohen_86}. Specifically, we modeled our mock galaxy with a spherically symmetric \citet{deVaucouleurs_59}-law profile, giving it an effective radius of $R_e = 14$~kpc, a surface brightness at $R_e$ of $\mu_V = 22.6$~mag~arcsec$^{-2}$, and a nominal PNe to luminosity ratio of $\alpha_{-2} = 20 \times 10^{-9}$~PNe~$L_{\odot}^{-1}$ \citep{Ciardullo+05, Ciardullo_10}. We then placed PNe randomly within the galaxy, using the assumption that the PNe were spatially distributed as the galaxy's light, and assigned the PNe absolute magnitudes drawn from the luminosity function given by equation~(\ref{eq:pnlf}) over the magnitude range between $M^*$ and $M^* + 6$.  We then placed the galaxy at a Virgo cluster-type distance modulus of $(m-M)_0 = 31.25$, and ``observed'' the system in a manner similar to that of the original Virgo cluster survey by \citet{Jacoby+90}, i.e., we adopted a field of view of $\sim 8\arcmin$, masked out the \OIII\ sources in the bright central regions of the galaxy (where all but the brightest PNe were difficult or impossible to detect), and assigned the remaining $m_{5007} \leq 27.2$ PNe to be part of a statistically complete sample. For this simple experiment, photometric errors were assumed to be negligible, and any two PNe separated by less than half the seeing FWHM (see \S\ref{sec:muse_parameters}) were considered to be unresolved and treated as a single source with the combined \OIII\ flux of the two objects.

We used our ``observed'' set of \OIII\ PN magnitudes to derive the distance to the mock galaxy in two different ways.  Our first analysis followed the methodology outlined in \citet{Ciardullo+89a} and simply fit the luminosity function of \OIII\ sources to equation~(\ref{eq:pnlf}) via the method of maximum likelihood.  Our second method took PN superpositions into account by applying the procedures described in \S\ref{sec:theory}.  

The results of 1,000 simulations show that there is no systematic offset between the distances derived via the approach described in this paper and the actual distance of the mock galaxy:  the most-likely distance modulus found for the galaxy is $(m-M)_0 = 31.25$ and the distribution of solutions is symmetrical about the mode.  The same cannot be said for the traditional method of analysis. 

Figure~\ref{fig:blend_simulation}  plots the difference between the true distance modulus of our mock galaxy and that found by simply fitting the luminosity function to a model function without taking blends into account.  Each curve represents the summed result of 10,000 simulations, where for each mock set of PN magnitudes, two posterior PDFs were computed:   one where all the PNe were assumed to be spatially resolved, and another where superposed PNe were treated as one source with the combined \OIII\ flux of both objects.  The curves show the difference between the two PDFs, transformed into a cumulative function.  (In other words, the curves give the probability of deriving a distance modulus at least $\Delta (m-M)_0$ smaller than the true value, due to the action of PN blends.)  The functions behave as expected.  According to the left-hand panel, if one excludes data from the central $40\arcsec$\ of the galaxy (where the surface brightness is high and PN detections are difficult), then, under $1\farcs 0$ seeing conditions, there is an $\sim 8\%$ chance that a distance derived from the PNLF method will be too small by at least 0.1~mag, and a $\sim 5\%$ chance that the systematic error will be at least $\sim 0.2$~mag.  Or, put another way, if our Virgo galaxy were observed in $1\farcs 0$ seeing, then, even if the central $40\arcsec$\ were excluded from the analysis, $\sim 55\%$ of the time superpositions would boost the magnitude at least one of the observed PNe by more than 0.05~mag, and $\sim 40\%$ of the time, this boosting would be more than 0.2~mag.

Obviously, these numbers increase rapidly as the image quality degrades.  But even under excellent ($0\farcs 5$) seeing conditions, our model with a $40\arcsec$\ exclusion radius predicts that 20\% of the time at least one of the observed PNe would have a magnitude at least 0.05~brighter than it should due to superpositions, and 3\% of the time, the galaxy's distance would be underestimated by 0.1~mag.  While one can mitigate the effect of superpositions by restricting PN surveys to the outer, lower-surface brightness regions of a galaxy, such a strategy has the obvious side effect of reducing the number of PNe available for study.   Since a precision PNLF distance requires $\sim 30$ to 50 PNe in the top $\sim 1$~mag of the PNLF, such an approach is not an option for lower luminosity galaxies or surveys that are restricted to a small field-of-view.  Moreover, as equation~(\ref{eq:lambda}) demonstrates, increased distance and poorer image quality are mathematically equivalent.  Thus, as the distance to a galaxy is increased, the effect of superpositions becomes more important. Any analysis which does not take image blending into account must therefore exclude more of a galaxy's bright, inner regions where the bulk of the PNe reside.

\null

\section{Analysis of the MUSE Hyperparameters}
\label{sec:muse_parameters}
As pointed out above, the MUSE integral-field unit spectrograph offers a number of advantages for PNLF studies, including its wide wavelength coverage, which allows interloping emission-line objects to be excluded via their line ratios, and a spectral resolution capable of disentangling superposed objects based on their differing radial velocities. But one disadvantage is MUSE's relatively small field-of-view, which, in its wide-field mode, is only $1\arcmin \times 1\arcmin$.  This constraint, coupled with the instrument's narrow effective bandpass, encourages PN surveys to be directed at the bright, inner regions of galaxies.  However, in these high surface-brightness regions, PN superpositions will be relatively common, and may compromise PNLF distance measurements.

\begin{figure*}
\includegraphics[width=18cm]{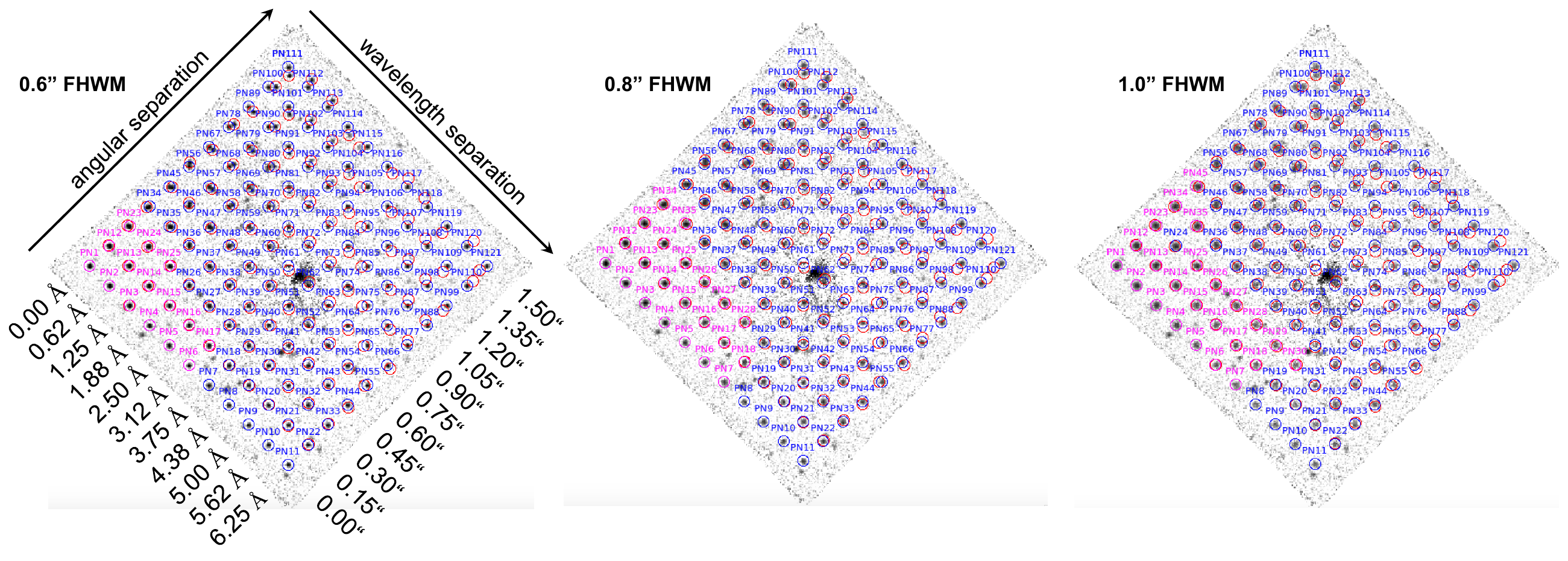}
\caption{The 15 continuum-subtracted data cube layers centered on redshifted \OIII\ $\lambda 5007$ for the 3600~sec MUSE observation of the central regions of NGC\,1380. Pairs of randomly oriented artificial PNe with $m_{5007} = 27.0$ have been inserted onto the cube with spatial and spectral separations that are given by the pairs' location in the grid.  For comparison, the limiting magnitude for completeness in $0\farcs 75$ seeing is $m_{5007} \sim 27.8$.  From left to right, the panels show simulations with $0\farcs 6$, $0\farcs 8$, and $1\farcs 0$ FWHM seeing. The PN pairs that are resolved into two separate objects are denoted in blue;  the magenta labels show pairs where the two objects cannot be distinguished.  Virtually all PNe pairs with angular separations greater than half the seeing FWHM or spectral separations greater than 3.75~\AA\ are easily resolved.  For fainter objects or pairs with large ($\gtrsim 0.5$~mag) brightness ratios, these limits can move by $\sim 0\farcs 1$ and $\sim 1$~\AA\null.  Note that in the central $\sim 5\arcsec$ of the galaxy, even the DELF technique cannot remove all of the stellar continuum.
}
\label{fig:mockblend}
\end{figure*}

To examine this problem, we first need to determine how the recorded image quality of a MUSE data cube and the spectral resolution of the instrument relate to our ability to identify closely separated PNe.  From equation~(\ref{eq:fblend}), the likelihood of a given PN being a superposition of multiple sources depends on five quantities:  the distance to the galaxy, the ratio of PNe per unit galaxy luminosity, the galaxy surface brightness at the position of the PN, and the spectral and spatial resolution of the observation.  The first two quantities are intrinsic to the galaxy and are the variables that one generally solves for when fitting the PNLF\null.  The measurements of galactic surface brightness can either be obtained via prior broadband surface photometry or from the galaxy continuum recorded in the MUSE data cube. The last two hyperparameters are a function of the resolution of the instrument and the observing conditions, and must be computed before the analysis.  

For the former, the wide-field mode of MUSE has a resolution of roughly $R \sim 2000$ or $\sim 150$~km\,s$^{-1}$ at 5000\,\AA\ \citep{Bacon+10}. Even with a careful examination of the profiles of the \OIII\ emission lines, one should not expect to detect PNe blends with velocity separations much less than this number.  Moreover, one would also expect this limit to be larger for faint targets, such as the PNe of distant galaxies.  

Similarly, we should not expect to detect PNe blends when their angular separations are much less than the image quality of the observation. The seeing of a MUSE data cube is recorded in several ways.   First, data from a Differential Image Motion Monitor \citep[DIMM;][]{Kornilov+07} are logged at the beginning and end of each MUSE exposure. At the same time, the MUSE Slow Guiding System \citep[SGS;][]{Zins+14} uses field stars in the telescope's focal plane to measure each exposure's minimum, maximum, and median image quality, along with the standard deviation of the distribution.  Both these systems provide a good approximation to the seeing on a MUSE data cube, but for the best estimate of image quality, we use the data cubes themselves and derive the FWHM of point sources present in the field.  This quantity, which is determined while calculating the aperture correction applicable to PN photometry, is detailed in \citet{Roth+21} and is the most reliable measure of an observation's spatial resolution.

We next need to know how the quantities of resolution and image quality relate to the variables $\Delta v$ and $\theta$ in equations~(\ref{eq:fblend}) and (\ref{eq:lambda}).  To do this, we ran a series of tests where pairs of artificial PNe were placed within a data cube with various spatial and wavelength separations. In the spatial direction, the component separations ranged  between $0\farcs 0$ and $1\farcs 5$ (with random orientations), while in the wavelength direction, the PN offsets ran from 0 to 6.25~\AA\ (0 to 375~km~s$^{-1}$).  We examined each pair of PNe in the exact same manner as that for the real data, i.e., we stepped through the data cube with the DS9 visualization tool \citep{Joye+03} and ``blinked'' through the data cube's layers, allowing subtle changes in the spatial and wavelength domains to be detected.  We then quantified our ability to distinguish PN pairs as a function of image quality, velocity separation, magnitude difference, and signal-to-noise. While knowing the positions of the objects in the frame beforehand may seem problematic, the goal of this experiment was not to test our ability to find \OIII\ sources, but to determine how reliably the two components of a blended pair of PNe could be disentangled.

Figure~\ref{fig:mockblend} shows one of these simulations, where pairs of equally bright ($m_{5007} = 27.0$) artificial PNe have been inserted into a 3600~second MUSE exposure of the Fornax lenticular galaxy NGC\,1380.  Each image shows the co-addition of 15 1.25~\AA\ slices of a continuum-subtracted data cube \citep[see][]{Roth+21}, with the spatial and spectral separation of each pair of \OIII\ $\lambda 5007$ emission lines denoted by its position in a grid.  The three images represent simulations with seeings of $0\farcs 6$, $0\farcs 8$, and $1\farcs 0$.  For all sources, both PNe are circled at their inserted locations. Sources where both components of the superposed PNe are detectable by our blinking procedure are labeled in blue (with circles of blue and red), while the unresolved pairs are shown in magenta (with both circles in magenta). Even a cursory inspection of the figure shows that PNe pairs with angular separations greater than half the seeing FWHM are almost always resolved, even when the objects have exactly the same radial velocity. Correspondingly, when the wavelength separation of the two emission lines is greater than $\sim 3$ pixels (3.75~\AA), the lines can be distinguished, even when the two PNe are in perfect spatial alignment.   At lower signal-to-noise and larger magnitude differences, the division between regions where the PNe can and cannot be resolved becomes slightly blurred.  For example, for bright objects, our ability to resolve two PNe with the same radial velocity is slightly better than the nominal value of half the seeing FWHM, while for faint PNe or PNe pairs with large ($\gtrsim 0.5$~mag) differences in brightness, the limit is slightly worse. However, this region of ambiguity is very narrow, roughly $0\farcs 1$ in the spatial direction and $\sim 1$~\AA\ in the spectral direction. Thus, to first order, we can assign $\theta$ to be half our measured value of the seeing FWHM, and $\Delta v = 200$\,km~s$^{-1}$.

\section{The Test Case of NGC 1380}
\label{sec:muse_testcase}
To illustrate the effect of PN superpositions on the analysis of MUSE data, we reanalyzed the data cube of the Fornax lenticular galaxy NGC\,1380.  The PNLF of this galaxy has been published multiple times -- first from a narrow-band imaging survey by \citet{Feldmeier+07}, then from the MUSE observations of \citet{Spriggs+20, Spriggs+21}, and most recently, from a reanalysis of the \citet{Spriggs+20} data cube using differential emission-line filtering \citep[DELF;][]{Roth+21}.  While this dataset may not be ideal for determining an absolute distance to the galaxy -- such a measurement requires  careful standard star observations and the presence of at least one bright point-source in the field for the photometric aperture correction -- it is more than sufficient for a differential analysis of the data.  Here we use the \citet{Roth+21} set of PN measurements, as their DELF technique delivers the highest signal-to-noise possible for the data and reaches the deepest limiting magnitude.

\begin{figure}
\centering
\includegraphics[width=3.35in]{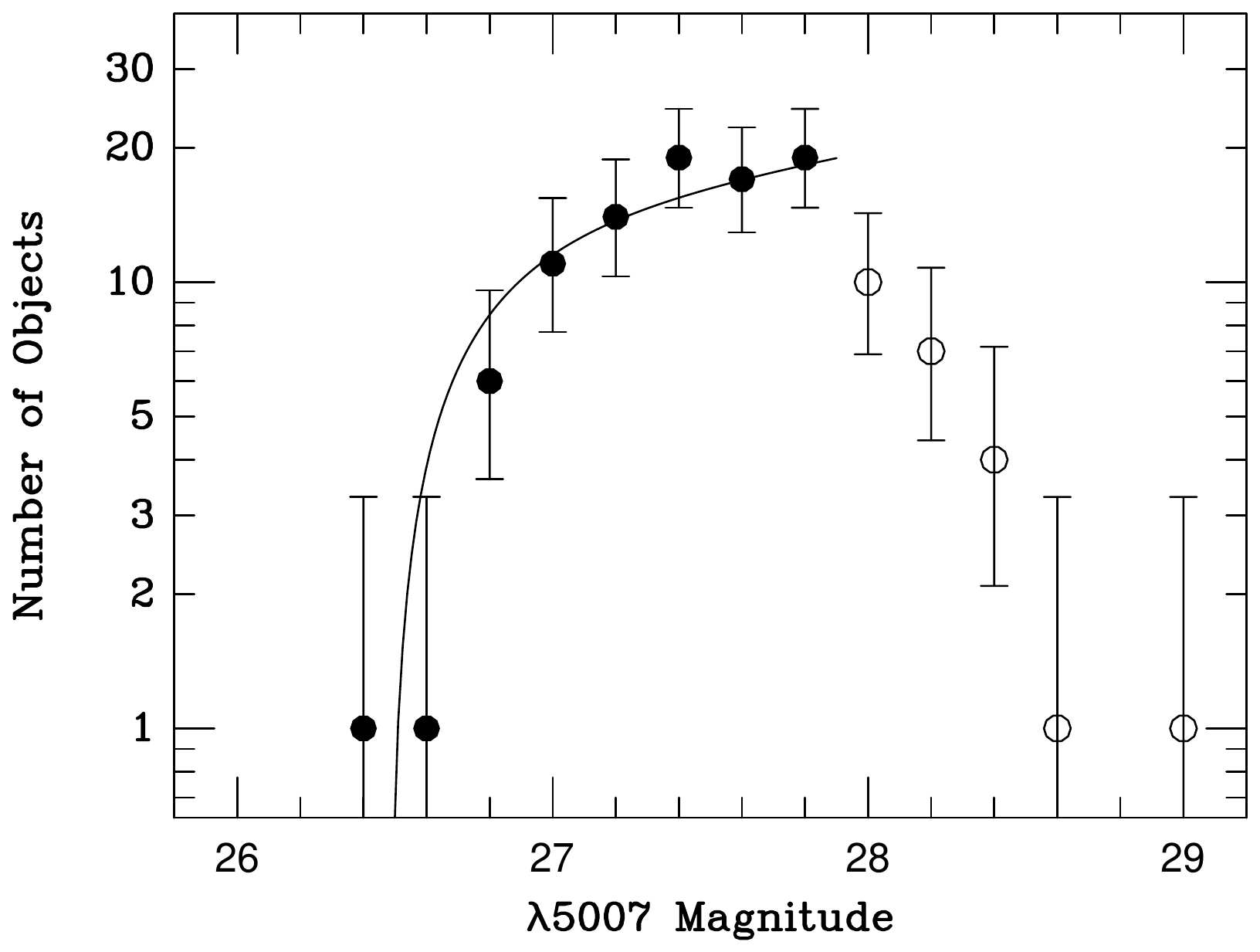}
\caption{The observed PNLF of the central MUSE field of NGC\,1380, binned into 0.2~mag intervals.  The error bars represent the $1\sigma$ confidence intervals of each bin \citep[see][]{Gehrels86}; the curve shows the most-likely fit.  The brightest PN is 0.25~mag more luminous than the second brightest object, but given the relatively low value of $\mu_V$ at its position, the likelihood of it having a magnitude brighter than $M^*$ due to a superposition is relatively low.}
\label{fig:ngc1380_pnlf}
\end{figure}

Figure~\ref{fig:ngc1380_pnlf} displays the PNLF of NGC\,1380 binned into 0.2~mag intervals.  The most notable feature of the PNLF is the presence of an \OIII\ source that is $\sim 0.25$~mag brighter than any other PN candidate in the galaxy.  This object has a MUSE spectrum that is perfectly consistent with that expected from a bright PN, with a very high \OIII/H$\beta$ ratio, negligibly faint lines of \NII\  and \SII\ \citep{Kreckel+17}, no observable continuum, an \OIII/H$\alpha$ ratio that falls within the locus for normal \OIII-bright PNe \citep{Herrmann+08},  and no evidence for velocity substructure in the \OIII\ $\lambda 5007$ line profile.  Yet because of its anomalously high line flux, both \citet{Spriggs+20} and \cite{Roth+21} excluded the object from their fits.  Our methodology allows us to check whether the existence of this source is consistent with the superposition hypothesis. We therefore keep the object in our statistical sample and include it in our analysis.

To facilitate our test, we restrict our dataset to PNe found within the single MUSE pointing on NGC\,1380's central regions. This removes one possible source of systematic error -- that associated with the slightly different photometric zero-points of the three different MUSE data cubes located at different positions in the galaxy.  In addition, because PN detections are difficult in regions with a bright, rapidly varying background, we mask out an $11 \farcs 0 \times 7\farcs 5$ elliptical isophote centered on the galaxy's nucleus.  This eliminates only 1~PN from the dataset, and ensures that the remaining sample of 78 PNe is reasonably complete to $m_{5007} = 27.8$.  If we sum up all the MUSE continuum measurements outside this ellipse, then the total amount of $V$-band light contained in the field is $m_V \sim 11.5$.

\begin{figure*}
\includegraphics[width=18cm]{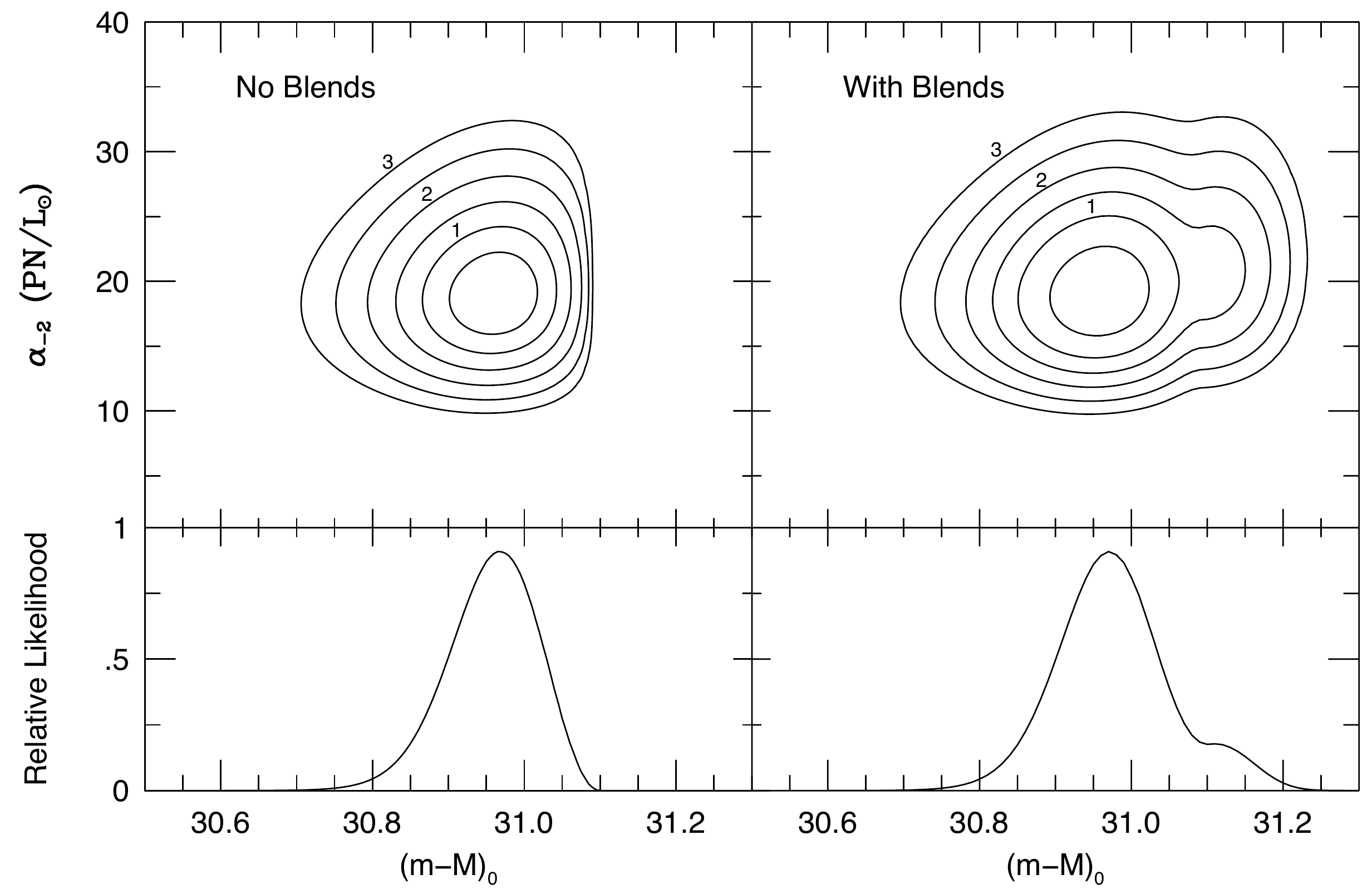}
\caption{The top panels show the likelihood contours for PNLF fits to the central MUSE field of NGC\,1380.  The abscissa is the dereddened distance modulus assuming $E(B-V) = 0.015$ \citep{Schlafly+11}; the ordinate is the number of planetaries with $M_{5007} < -2$ normalized to the amount of bolometric galaxy luminosity sampled.  The contours have been drawn at $0.5 \sigma$ intervals and are labeled. The bottoms panels marginalize over $\alpha_{-2}$ and give the relative likelihood versus distance modulus.  The panels of the left show a conventional fit to equation~(\ref{eq:pnlf}); the right panels show the solution which includes the possibility of PN superpositions.   In this case, the surface brightness upon which the brightest PN is superposed is relatively low, hence the most-likely distance modulus is relatively unaffected.  However, the more modern code admits the possibility that the brightest PN in the sample is a superposition of two objects, and shows likelihood at larger distances. 
}
\label{fig:contours}
\end{figure*}

To perform our analysis, we first used the MUSE data cube to measure the galaxy's continuum at the location of each PN (i.e., the values of $\mu_V$ in equation~(\ref{eq:lambda})), and, to estimate the $\sigma_{\rm gal}$ values appropriate for the PNe, we referred to the line-of-sight velocity dispersion map published by \citet{Iodice+19}. In general, we expect most spatial superpositions to be unresolved by MUSE, as the galaxy's velocity dispersion is relatively small, dropping from $\sim 190$~km~s$^{-1}$ at our inner isophote for PN detections to $\sim 120$~km$^{-1}$ at the edge of the field.  The values of $\mu_V$ and $\sigma_{\rm gal}$ then allowed us to derive the distance to the galaxy with and without the inclusion of superpositions in the PNLF analysis.

Figure~\ref{fig:contours} shows the likelihood contours for the fits to NGC\,1380's PNLF, with the left panel giving the solution from a simple matching to equation~(\ref{eq:pnlf}) via the method detailed in \citet{Ciardullo+89a}, and the right panel found from the methodology of \S\ref{sec:theory}.  From the figure, it is clear that both solutions yield the same \textit{most-likely} distance modulus, $(m-M)_0 \approx 30.97$ \citep[assuming a foreground extinction of $E(B-V) = 0.015$;][]{Schlafly+11}. This is not unexpected, as the shape of the galaxy's PNLF cutoff is extremely well defined by the $\sim 50$~PNe in the top $\sim 1$~mag of the luminosity function. But in a conventional analysis, NGC\,1380’s distance modulus cannot be greater than $(m-M)_0 \approx 31.1$, as a larger distance would imply an \OIII\ $\lambda 5007$ luminosity for the brightest PN that is greater than $M^*$. This hard limit is reflected by the shape of the likelihood contours in the left-hand panel.

The possibility of PN superpositions softens this constraint.  In the case of NGC\,1380, the galaxy's most luminous PN is located on a region of relatively low surface brightness ($\mu_V \sim 19.0$~mag~arcsec$^{-2}$); this fact, along with the PN survey's very good ($0\farcs 75$) image quality, means that the likelihood of that PN being a superposition of two objects is relatively low, $\sim 2\%$. This improbability is reflected by the error contours in the right panel of Figure~\ref{fig:contours}, where the distortion of the likelihood contours only becomes apparent beyond the 68\% confidence level.  Still, the effect is non-negligible, as it causes the positive-side error bar associated with the distance to increase from 0.05~mag to 0.08~mag.  This bias would not be recognized by conventional analyses. 

The differences in the shape of the two PDFs are important.  The likelihood contours for NGC\,1380 are quite robust, due to the fact that the data extend more than a magnitude down the PNLF and there are over 80 PNe populating this region.  It is easy to imagine that at the larger distances needed to investigate $H_0$, less of the PN luminosity function will be surveyed, and fewer objects will be detected.  Furthermore, the chance of PN superpositions will be increased, leading to greater distortions in the PNLF and more likelihood at larger distances.

Figure~\ref{fig:contours} also illustrates the problem of only using most-likely distances in a Hubble Constant analysis.  While this value may be best for a single galaxy, its use for an ensemble of objects biases the overall solution towards distances that are too small and Hubble Constants that are too large.  In this era of precision cosmology, even small systematics such as this cannot be tolerated. 

Finally, we note that the shape of NGC\,1380's likelihood contours is both asymmetrical and distorted towards greater distances.  This is due to the shape of the model PNLF being fit:  due to the contribution of superpositions, this function now has two regions of sharp declines, one for the single object component and one produced by superpositions (see Figure~\ref{fig:pnlf_blend}).  A cross-correlation between the model and the observed PNLF can therefore produce two peaks, one associated with PNLF's shape, and the other derived from the superposition ``echo'' at brighter absolute magnitudes.  In an extreme case with a sparsely-populated luminosity function that extends only a few tenths of a magnitude beyond $M^*$, it is possible for the likelihood function to become bimodal.  Fortunately, for the large galaxies often targeted by the PNLF, this should usually not be an issue.

\section{Discussion}
\label{sec:discussion}

Qualitatively, there are three types of PN superpositions which occur within a galaxy.  The first is the case where the summed \OIII\ flux of two objects is fainter than $m^*$ by more than $\sim 0.5$~mag.  When this happens, the existence of the superposition has a negligible effect on the shape of the PNLF and can safely be ignored.  The second is when the resultant blend produces an object significantly ($\gtrsim 0.25$~mag) brighter than $m^*$.  When this occurs, it is obvious from even a cursory inspection that the source's \OIII\ luminosity is inconsistent with the presumed shape of the PNLF (equation~\ref{eq:pnlf}) and the object is excluded from the analysis.  Although this ad hoc deletion is not completely satisfying, it does minimize the effect of a single anomalous object on the derived distance to a galaxy.

The third case is the one that is hardest to address -- when the summed flux of two merged PNe is within $\sim 0.25$ of $m^*$.  A visual inspection of the luminosity function may, or may not identify these objects as ``overluminous,'' and they will often be included in the sample of PNe being fit.  Yet because of the unique shape of the PNLF, such objects will have an outsized effect on a system's derived distance, especially if the survey depth is relatively shallow or the bright-end of the PNLF is sparsely populated.  The result could be a systematic error, such as that described by \citet{Ferrarese+00}.  The procedure described in \S\ref{sec:theory} addresses this issue, as it folds possible superpositions in the PNLF analysis.

Of course, as shown in equation~(\ref{eq:lambda}), the likelihood of a superposition is a strong function of galaxy surface brightness, distance, and seeing. In galactic halos or in the disks of nearby systems, the chance of two PNe falling onto the same detector element is low; in these cases, the distance to the system can be found simply by using a standard $\chi^2$ or maximum-likelihood statistic.  But in the inner regions of galaxies beyond $\sim 10$~Mpc, superpositions become more prevalent. To estimate how prevalent, one can simply use the expression for expectation value given by equation~(\ref{eq:lambda}) and calculate
\begin{equation}
    P = \prod_{i=1}^N w_{1,i} 
\end{equation}
where the $w_{1,i}$ values represent the likelihood that each PN is composed of only one object (see equation~\ref{eq:weight}), and $N$ is the total number \OIII\ sources in sample.  If this number is more than a few percent, then the methodology presented in this paper is likely needed.

Finally, it is important to note that PN superpositions may not be the sole cause of the PNLF's breakdown at large distance. Indeed, as mentioned above, the most luminous PN candidate in our test-case galaxy (NGC\,1380) is unlikely to be a superposition, and several other ``overluminous'' PNe are known to exist in regions of relatively low galaxy surface brightness (see, for example, \citet{Spriggs+21} and \citet{Scheuermann+22}). This suggests that other factors may be responsible for ``overluminous PN'' phenomenon, including the mis-classification of objects such as supernova remnants, Wolf-Rayet nebulae, and background Ly$\alpha$-emitting galaxies \citep[e.g.,][]{Kreckel+17, Kudritzki+00, Longobardi+15}, the inclusion of foreground intracluster stars in the PN sample \citep[e.g.,][]{Feldmeier+04, Doherty+09, Longobardi+13}, and/or an incorrect representation of the true shape of the PNLF's bright-end cutoff \citep{Mendez+08, Valenzuela+19}.  MUSE's wide wavelength coverage and $R \sim 2000$ spectral resolution address the first issue:  by examining the line profile of the putative \OIII\ $\lambda 5007$ line and various line ratios (such as \OIII/H$\alpha$, H$\alpha$/\NII\ and H$\alpha$/\SII, objects such as \HII\ regions, supernova remnants, and background galaxies can reliably be excluded from the sample of PN candidates \citep{Trainor+15, Herrmann+08, Kreckel+17}.  The second issue applies only to galaxies in clusters, and can also be addressed (albeit with difficulty) via the velocity resolution provided by the MUSE spectrograph \citep{Arnaboldi+22}.   Thus, the greatest issues now facing the use of PNe for precision distance determinations is the exact shape of the PNLF's bright-end cutoff, whether that shape changes with stellar population, and whether there is a separate class of objects with the same spectral signature as PNe.  Since there is, as yet, no robust theory for the behavior of the PNLF \citep[see the discussion in][]{Ciardullo_22}, this question can only by addressed by continuing to accumulate high-quality data in a variety of stellar systems and using extreme value theory to identify outliers \citep{Coles_01}.

\acknowledgments
This work was supported by the NSF through grant AST2206090 and has made use of the NASA/IPAC Extragalactic Database (NED), which is funded by the National Aeronautics and Space Administration and operated by the California Institute of Technology.  The Institute for Gravitation and the Cosmos is supported by the Eberly College of Science and the Office of the Senior Vice President for Research at the Pennsylvania State University.  We thank the anonymous referee for valuable suggestions to improve the presentation.

\newpage
 
\bibliography{pagb.bib}

\end{document}